# Identifying vital nodes by Achlioptas process

Zhihao Qiu[1], Tianlong Fan[2,1,3], Ming Li[4] and Linyuan Lü[1,3,5,*]

[1] Institute of Fundamental and Frontier Sciences, University of Electronic Science and Technology of China, Chengdu 611731, China.

[2] Department of Physics, University of Fribourg, Fribourg 1700, Switzerland

[3] Yangtze Delta Region Institute (Huzhou), University of Electronic Science and Technology of China, Huzhou 313001, P. R. China.

[4] Department of Thermal Science and Energy Engineering, University of Science and Technology of China, Hefei 230026, People's Republic of China

[5] Complex Systems Lab, Beijing Computational Science Research Center, Beijing 100193, People's Republic of China.

[*] Author to whom any correspondence should be addressed.
E-mail: linyuan.lv@uestc.edu.cn (L. Lü)

**Abstract**

The vital nodes are the ones that play an important role in the organization of network structure or the dynamical behaviours of networked systems. Previous studies usually applied the node centralities to quantify the importance of nodes. Realizing that the percolation clusters are dominated by local connections in the subcritical phase and by global connections in the supercritical phase, in this paper we propose a new method to identify the vital nodes via a competitive percolation process that is based on an Achlioptas process. Compared with the existing node centrality indices, the new method performs overall better in identifying the vital nodes that maintain network connectivity and facilitate network synchronization when considering different network structure characteristics, such as link density, degree distribution, assortativity, and clustering. We also find that our method is more tolerant of noisy data and missing data. More importantly, compared with the unique ranking list of nodes given by most centrality methods, the randomness of the percolation process expands the possibility space of the optimal solutions, which is of great significance in practical applications.

Keywords: complex networks, vital nodes identification, network percolation, Achlioptas process

## 1. Introduction

Identifying and ranking vital nodes in complex networks is one of the important research fields in network science [1–4]. By identifying and controlling vital nodes, one can facilitate the spread of information [5] and promote products [6,7], inhibit the spread of the epidemic [8,9], prevent the cascading failures [10–12] and better control the network [10–12] and so on. According to the difference in principle of methods, Ref. [2] divided representative vital nodes identification methods into four categories: (1) structural centralities, such as degree centrality [13] and coreness [14]; (2) iterative refinement centralities, such as eigenvector centrality [15]; (3) node operation, such as connection-sensitive method [16]; (4) dynamics-sensitive methods, such as dynamics-sensitive centrality [17]. Besides the methods in these four categories, there is another group of methods that considering the network percolation process. Piraveenan et al. [18] first proposed percolation centrality that quantifies the relative impact of nodes based on their topological connectivity, as well as their percolation states, which can be considered as a variant of the betweenness centrality with weighting based on percolated state. Recently, Morone et al. [19] pointed out that the problem of identifying vital nodes can be exactly mapped onto the so-called optimal percolation. Besides, there are many methods defined in terms of the percolation process, such as articulation point-targeted attack strategy, greedy articulation points removal strategy [20], simulated annealing [21], and the method based on percolation critical state [5].

In general, percolation considers the behavior of the clusters formed by the occupied nodes/links of a network,

which can be realized by inserting the corresponding nodes/links into the network according to its structure [22, 23]. Compared with the traditional indices of node importance, the indices extracted from the percolation process have their unique advantages. First, the clusters formed in the percolation process have universal properties, which do not depend on the specific network attributes, such as the network scale and average degree. Second, the percolation process not only reflects the structural features of networks but also involves the characteristics of dynamics, so it can expose the influence of nodes on both structure and dynamics of networks. Besides, the randomness of percolation process allows the existence of fluctuation on the ranking list and thus may yield multiple solutions. Note that not all the methods defined in terms of percolation have this advantage, for example, k-core decomposition [14], optimal percolation [19], history-dependent percolation [24] and greedy articulation points removal [20]. And systematic reviews on the network percolation process can be found in Refs. [25] and [26]. A form of explosive percolation (EP) is realized by applying a competitive rule for edge addition via what is known as an Achlioptas Process [27-30], which suppresses the growth of the largest cluster when nodes are activated one by one. In the early stage of this percolation process, the size of the largest cluster grows slowly, but once the number of activated nodes approaches a certain threshold, the size of the largest cluster will grow rapidly and a giant connected component scaling in the order of system size will emerge [29]. This suggests that the later the nodes are activated, the more important they are to the network both structurally or functionally.

In this paper, we propose a universal method of vital nodes identification based on the explosive percolation, which covers all the above advantages of percolation-type methods. Explosive percolation and similar competitive mechanisms are often used to describe the growth of networks [31], here we utilize the growth process to identify vital nodes, i.e. the later growing node which also is seen as the winner of the competitive process are more likely to be the vital nodes. The analytical results show that the nodes identified by this method have a more important impact on the connectivity and synchronization dynamics of networks than other classical indices. Then, this paper further analyzes the influence of network structure on the performance of the new method under different network features, including link density, degree distribution, assortativity, clustering feature, and also explores the influence of human intervention (i.e. node protection strategy) and non-structure factor (i.e. containing missing or noisy data) on the robustness of the method. The results show that the new method can maintain good performance under different conditions. Then we demonstrate a practical advantage of the new method, namely how it can be used to identify alternative rankings that achieve similar performance. Finally, we discuss its computational complexity problem.

## 2. Results

### 2.1 Explosive Percolation (EP) method

We propose a new method, named the Explosive Percolation (EP) method, that utilizes the Achlioptas process as described below to quantify nodes importance. Given a simple undirected network $G(V, E)$, where $V$ and $E$ represent nodes and links, respectively. We denote the state of node $i$ by $\sigma_i$. Initially, $\sigma = 0$ is assigned to all the nodes meaning they are inactive. Then the nodes are activated ($\sigma = 1$) one by one according to the following rules:

(a) Two nodes $j$ and $k$ are chosen randomly from all the inactive nodes. Assuming nodes $j$ and $k$ are activated respectively, then their respective cluster sizes are

$$\xi_j = \sum_{i \in J} \sigma_i, \xi_k = \sum_{i \in K} \sigma_i, \quad (1)$$

where $J$ ($K$) is the set of nodes containing all connected active nodes in the cluster to which the node $j$ ($k$) belongs.

(b) Activate the node with the smaller cluster size $\xi$, and the other one remains inactive, namely

$$\begin{cases} \sigma_j = 1, \sigma_k = 0, \; if \; \xi_j < \xi_k \\ \sigma_j = 0, \sigma_k = 1, \; if \; \xi_j > \xi_k \end{cases} \quad (2)$$

If $\xi_j = \xi_k$, one of the two nodes will be randomly activated.

(c) Repeat (a) and (b) until all the nodes in the network are activated. The order of the node activation of node $i$ is denoted by $r_i$. Therefore, the later the node is activated, the more important it is.

Because the two potential nodes are randomly selected from those not yet activated, the average value of multiple realizations is used to reduce the fluctuation, and thus the final importance of node $i$ is defined as

$$\varphi_i = \frac{1}{w} \sum_{t=1}^{w} r_i^t \quad (3)$$

where $r_i^t$ is the activation order of node $i$ in the $t$-th realization, $w$ is the number of realizations. In this paper, we set $w = 500$. The larger the value $\varphi_i$, the more important the node $i$.

### 2.2 Empirical Data

The performance of our method, or called EP method, is investigated on six disparate empirical networks [32], including animal social network of zebra (Zebra), the collaboration network of jazz musicians (Jazz), the US air transportation network (USAir), the collaboration network of scientists working on network science (NS), the email communication network of the University at Rovira i Virgili in Spain (Email), and the US power grid network (Powergrid). The basic statistical properties of the six networks are shown in Table. 1.

## 2.3 Performance Evaluation of EP method

*2.3.1 Network dismantling* Network dismantling, or network attack [33,34], is destroying network structure and function by disabling nodes or links. The efficiency of network dismantling can be used to assess the performance of vital node identification methods. Specifically, a method performs better than others if the network collapses faster by removing nodes one by one according to the ranking given by this method. For comparison, we also consider some well-known methods, including Degree centrality (DC) [1] (i.e., the number of direct neighbors), Closeness centrality (CC) [35,36] and Eigenvector centrality (EC) [2,15,37], Collective Influence (CI) [19,38] and Adaptive Degree (AD) [38,39].

To quantitatively evaluate the performance of the method, a metric, called Robustness [2,40–42], is defined as node number N, link number M, average degree $<k>$, clustering coefficient C, and assortativity coefficient r.

**Table 1.** The basic statistical properties of the six empirical networks.

| Network | N | M | $<k>$ | C | r |
|---|---|---|---|---|---|
| Zebra | 27 | 111 | 8.222 | 0.8759 | 0.7177 |
| Jazz | 198 | 2742 | 27.69 | 0.6174 | 0.0202 |
| USAir | 332 | 2126 | 12.8 | 0.6252 | -0.2078 |
| NS | 379 | 914 | 4.823 | 0.7412 | -0.0816 |
| Email | 1133 | 5452 | 9.624 | 0.2201 | 0.0782 |
| Powergrid | 4941 | 6594 | 2.669 | 0.0801 | 0.0034 |

$$R = \frac{1}{N}\sum_{Q=1}^{N} \Omega(Q). \quad (4)$$

Here, $\Omega(Q)$ is the normalized value of the largest cluster size after removing the first $Q$ nodes, and the normalization factor $1/N$ allows the Robustness values of networks of different sizes to be compared [40]. Note that the vital nodes identified by different methods are the first to be removed. Clearly, a smaller $R$ corresponds to a quicker decrease of $\Omega(Q)$ and better attack effect.

Fig. 1 shows the changes of the normalized size of the largest cluster of the six empirical networks when nodes are removed following the order ranked by different methods. The results generally show that EP method makes all these networks dismantled more rapidly. Notably, in Fig. 1(b), since Jazz is a dense network with high average degree, it is difficult to broken down into small clusters at the early stage of the dismantling procedure, leading to a worse performance initially. The Robustness values of the six indices on six networks are shown in Table 2, where in each line the best performing index is in bold, with EP performing best in all cases. This indicates that EP method has good performance in dismantling network. In other words, the vital nodes identified by EP method are more important for maintaining network connectivity. This is mainly because some specific structures of the network may limit the efficacy of other methods if only consider either local or global information, while EP method captures both local and global information of the network. For instance, the degree centrality may underestimate the importance of bridge nodes (i.e., nodes that connect two clusters) with low degrees.

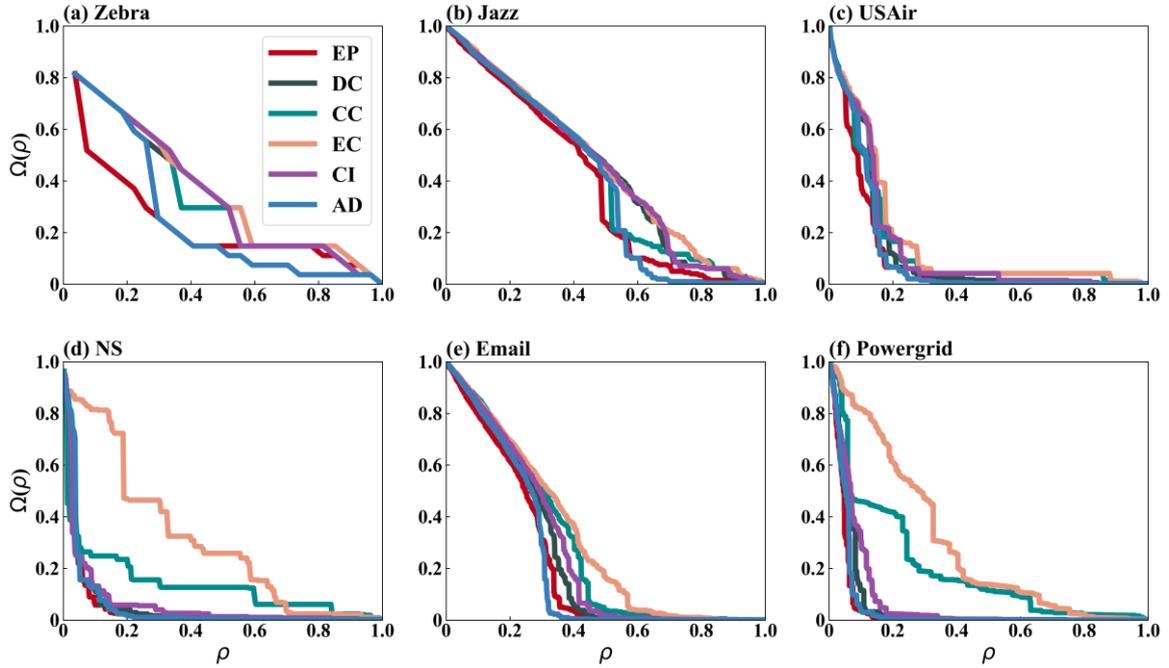

**Figure 1.** The performance of six indices investigated via network attack. Where $\rho=Q/N$ is the ratio of removed nodes, $\Omega(\rho)$ is the relative size of the largest cluster size after removing $\rho N$ nodes.

**Table 2.** Robustness $R$ of six indices for six empirical networks.

| Network | EP | DC | CC | EC | CI | AD |
|---|---|---|---|---|---|---|
| Zebra | **0.2235** | 0.3388 | 0.3347 | 0.3525 | 0.3443 | 0.2496 |
| Jazz | **0.3802** | 0.4407 | 0.4194 | 0.4549 | 0.4438 | 0.3912 |
| USAir | **0.1014** | 0.1256 | 0.1439 | 0.1637 | 0.1395 | 0.1076 |
| NS | **0.0465** | 0.0539 | 0.1335 | 0.3083 | 0.0610 | 0.0507 |
| Email | **0.2236** | 0.2518 | 0.2893 | 0.3163 | 0.2691 | 0.2237 |
| Powergrid | **0.0464** | 0.0615 | 0.1972 | 0.2938 | 0.0786 | 0.0510 |

*2.3.2 Pinning control.* Pinning control is an effective way for facilitating synchronization [43,44] which means that all nodes in the whole system reach the same state within a finite time by selectively pinning a fraction of nodes in the system. Specifically, the state of each node in the initial system is given randomly, and pinning a node means leading it to reach the target state by exerting a control action. With feedback control, the other nodes will reach the same target state after a fraction of nodes are pinned. We next verify the validity of vital nodes identified by EP method in synchronization dynamics.

For a simple connected network $G(V, E)$ consisting of $N$ linearly and diffusively coupled nodes, the system state equation is

$$\dot{x}_i = f(x_i) + \theta \sum_{j=1}^{N} l_{ij} \Gamma(x_j) + U_i(x_1, \dots, x_N) \quad (5)$$

where $x_i \in R^n$ is the state vector, $f(\cdot)$ denote the self-dynamics of node $i$, $\theta$ is coupling strength, $\Gamma(x_j): R^n \to R^n$ the inner coupling matrix connecting different components of a state vector. $U_i$ describes the control on node $i$, $L = [l_{ij}]_{N \times N}$ is called the Laplacian matrix [45] of the network, which satisfies the following conditions, if $(i,j) \in E$, then $l_{ij} = -1$; if $(i,j) \notin E$, and $i \neq j$, then $l_{ij} = 0$; if $i = j$, then $l_{ii} = -\sum_{j \neq i} l_{ij}$.

We successively pin nodes one by one according to the given ranking and quantify the synchronizability of the pinned networks by the reciprocal of the smallest nonzero eigenvalue of the principal submatrix [46] of the network, namely $1/\mu_1(L_{-Q})$, where $Q$ is the number of pinned nodes, and $L_{-Q}$ describes the principal submatrix which is obtained by deleting the $Q$ rows and columns corresponding to the $Q$ pinned nodes from the original Laplacian matrix $L$. A smaller value of $1/\mu_1(L_{-Q})$ corresponds to a higher synchronizability of the network after pinning $Q$ nodes.

To compare EP method with other methods, we pin nodes successively corresponding to different centrality indices and calculate $1/\mu_1(L_{-Q})$. Similar to the definition of Robustness, we here introduce an indicator, named Pinning Efficiency [46], to evaluate the performance of different indices on pinning control, which is defined as

$$P = \frac{1}{Q_{max}} \sum_{Q=1}^{Q_{max}} \frac{1}{\mu_1(L_{-Q})} \quad (6)$$

where $Q_{max} = 0.3N$ is the maximum number of pinned nodes. A smaller $P$ means a higher synchronizability and thus a better performance of the pinning order according to an index. We compare EP method with DC, CC, EC, CI, and AD in pinning control and the results are shown in Fig. 2 and Table 3 where the best-performed index in each line is in bold. We can see that the EP method ranks first in two networks of the six and rank second in the other four networks, and thus performs overall the best along with AD.

**Table 3.** The Pinning Efficiency $P$ of six indices for six empirical networks.

| Network | EP | DC | CC | EC | CI | AD |
|---|---|---|---|---|---|---|
| Zebra | **1.257** | 2.666 | 2.626 | 2.859 | 2.823 | 2.469 |
| Jazz | **1.749** | 1.793 | 1.792 | 1.962 | 1.814 | 1.779 |
| USAir | 7.353 | 7.711 | 7.770 | 8.527 | 8.218 | **5.955** |
| NS | 13.568 | 15.681 | 36.546 | 30.400 | 19.768 | **12.673** |
| Email | 2.887 | **2.880** | 2.937 | 3.531 | 2.898 | 2.898 |
| Powergrid | 77.226 | 82.579 | 446.154 | 833.551 | 148.516 | **61.825** |

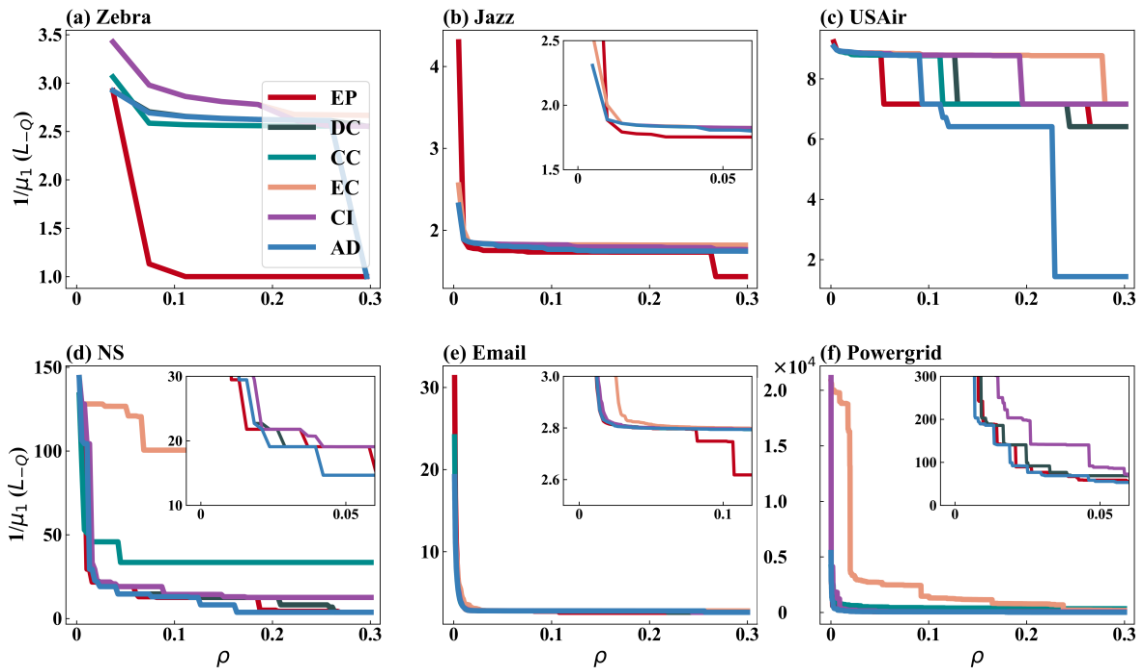

**Figure 2.** The performance of six indices investigated via the pinning control. $\rho = Q/N$ is the ratio of pinned nodes, $1/\mu_1(L_{-Q})$ is the reciprocal of the smallest nonzero eigenvalue of the principal submatrix of the network.

## 2.4 The influence of network features on the performance of EP method

The universality of a method on different networks is an important factor related to performance and practicability. This section studies the impact of different network statistical features on EP method performance, including link density, degree distribution, assortativity, and clustering.

### 2.4.1 Link density and degree distribution.
The average degree <k> of a network is used to describe the density of links, which is the most basic factor affecting the performance of methods. Here we constructed three Erdos-Renyi (ER) [47] networks with 1000 nodes and <k> of 5, 10, 15, respectively. Then, the performances of the four indices including EP, DC, CC, and EC are compared. As shown in Fig. 3a, a sparser network corresponds to better performance for all these indices, while EP method outperforms the other three in all cases. Fig. 3b presents the performance of four indices in three classic model networks, Erdos-Renyi network (ER) [47], Watts-Strogatz network (WS) [48], and Barabasi-Albert network (BA) [13], which feature the randomness, small-worldness, and scale-free property of connections, respectively. All the three networks have 1000 nodes and 5000 links. The link rewiring probability of WS is 0.1. The results showed in Fig. 3b clearly illustrate that in all the three model networks the EP method has the best performance.

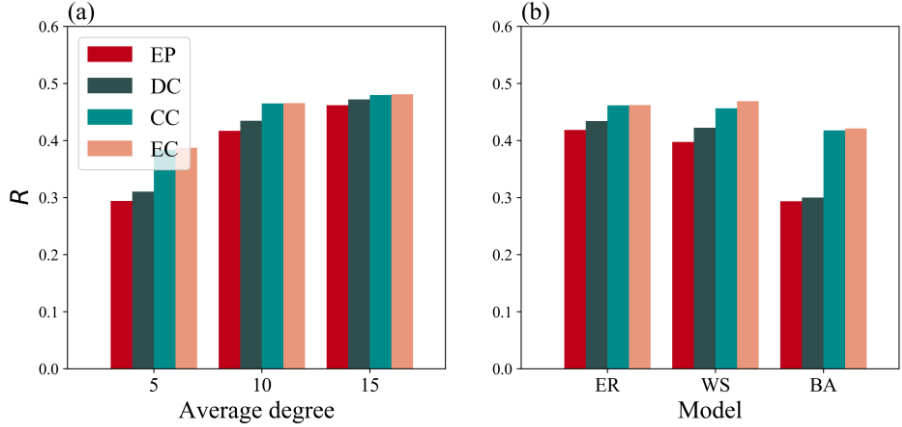

Figure 3. The influence of average degree and network models on the performance of the four indices in nodes attack. All the results are the average of 100 realizations.

*2.4.2 Assortativity.* Network assortative is another property that greatly affects method performance [3,49,50]. A network is assortative if nodes with a high degree tend to connect to other nodes with a high degree; on the contrary, the network is disassortative. Here, we calculate the degree of assortativity of a network [1,45,49] using the Pearson correlation coefficient [45,49,51,52], namely

$$r = \frac{M^{-1}\sum_{e_{ij}\in E(G)}(k_i k_j) - \left[M^{-1}\sum_{e_{ij}\in E(G)}\frac{1}{2}(k_i+k_j)\right]^2}{M^{-1}\sum_{e_{ij}\in E(G)}\frac{1}{2}(k_i^2+k_j^2) - \left[M^{-1}\sum_{e_{ij}\in E(G)}\frac{1}{2}(k_i+k_j)\right]^2} \quad (7)$$

where $M$ is the number of links in the network, $k_i$ and $k_j$ are the degree of node $i$ and node $j$, respectively, and the sum is over all the links. If $r = 1$, the network has the largest assortativity, while $r = -1$ indicates the largest disassortativity. The networks used in the experiments are constructed based on BA networks through the following reconnection rules [49]:
(a) Randomly select two links $e_{mn}$ and $e_{ij}$ without common nodes, then delete the two links.
(b) Rank nodes $m, n, i, j$ according to their degrees.
(c) To be assortative, connect the two nodes with the highest degrees, and then connect the remaining two nodes; to be disassortative, connect the two nodes with the highest and the lowest degrees, and then connect the remaining two nodes.
(d) Repeat steps (a)-(c) until the assortativity coefficient $r$ reaches the target value.

Robustness values for the six networks with different assortativity coefficients are listed in Table. 4, here A1-A3 denotes assortative networks while D1 and D2 are disassortative networks, respectively. In each line, the best-performed index is in bold, where all the networks have the same number of nodes $N = 1000$ and links $M = 4000$. According to the results, networks with lower assortativity are more easily collapse and EP method performs best in most networks. In fact, in the previous section, the assortativity of the three networks ER, WS, and BA decreases in turn, the Robustness of each index in these three networks also decreases in turn, which is consistent with the conclusion in Table. 4. This is because the nodes with large degrees are usually connected and form a network core in assortative networks, which makes the network robust against destruction; while the disassortative network is more tree-like and shows obvious vulnerability at the outset.

*2.4.3 Clustering coefficient.* The clustering coefficient [1] of a node assesses the density of the triangular pattern in its neighborhood. It can be also understood as the probability that two neighbors of the node are connected, namely

$$C_i = \frac{2\mu_i}{k_i(k_i-1)} \quad (8)$$

where $k_i$ is the degree of node $i$, $\mu_i$ is the number of links between the neighbors of node $i$. Moreover, the clustering coefficient of a network is the average clustering coefficient of all nodes in the network

$$C = \frac{1}{N}\sum_{i=1}^{N} C_i. \quad (9)$$

When increasing the clustering of a network with a fixed number of links, the local connections of the network become denser. However, the compact of the connections between neighbors comes at the expense of diluting other connections. Many previous studies have already shown that this makes the network easy to be destroyed [49,53–56], meaning that all the methods evaluated by the Robustness will perform better with the increasing clustering coefficient. The experiments are carried out on the scale-free networks (with the degree distribution follows $P(k)\sim k^{-3}$) with different clustering coefficients and the fixed number of nodes $N = 1000$ and links $M = 4000$. The networks are constructed by Holme and Kim algorithm [57]. The Robustness for six networks is shown in Table 5, here C1~C5 denotes networks with different clustering coefficients respectively, and the best-performed index is in bold. The results illustrate that EP method is superior to the other three indices in both high-clustering and low-clustering networks.

**Table 4.** Robustness $R$ for six networks with different assortativity. All the results are the average of 100 realizations.

| Network | $r$ | EP | DC | CC | EC |
|---|---|---|---|---|---|
| A1 | 0.35 | **0.4609** | 0.4810 | 0.4962 | 0.4956 |
| A2 | 0.17 | **0.3967** | 0.4301 | 0.4696 | 0.4736 |
| A3 | 0.05 | **0.3171** | 0.3370 | 0.4309 | 0.4378 |
| BA | -0.05 | **0.2539** | 0.2600 | 0.3887 | 0.3968 |
| D1 | -0.17 | **0.2238** | 0.2307 | 0.4096 | 0.4239 |
| D2 | -0.35 | 0.1140 | **0.1138** | 0.2701 | 0.4805 |

**Table 5.** Robustness $R$ for six networks with different clustering coefficient. All the results are the average of 100 realizations.

| Network | $C$ | EP | DC | CC | EC |
|---|---|---|---|---|---|
| BA | 0.03 | **0.2539** | 0.2600 | 0.3887 | 0.3968 |
| C1 | 0.07 | **0.2498** | 0.2560 | 0.3879 | 0.3945 |
| C2 | 0.15 | **0.2368** | 0.2426 | 0.3871 | 0.3915 |
| C3 | 0.24 | **0.2284** | 0.2328 | 0.3879 | 0.3894 |
| C4 | 0.32 | **0.2257** | 0.2312 | 0.3874 | 0.3866 |
| C5 | 0.46 | **0.2100** | 0.2194 | 0.3831 | 0.3782 |

## 2.5 The influence of non-structure factors on the performance of EP method

In practice, the performance of a method is not only affected by network structure, but also by some non-structure factors, such as protection strategy, or system noise, which are studied in this section.

*2.5.1 Protection strategy.* Due to the restrictions of objective conditions, sometimes the attack is unable to be applied to some nodes due to protection. To model these situations, $n$ nodes are randomly selected for protection, so that the attack on them is invalid. In Fig. 4, $\rho = n/N$, $n \in [0, 0.1N]$ denotes the fraction of protected nodes, where $N$ is the network size, $R(\rho)$ is the Robustness when protecting $\rho N$ nodes of the network. All the results are the average of 100 realizations. The results in Fig. 4 show that the Robustness of all four indices, indicating that the protection strategy degrades the performance of network attacks. However, we can find that the EP method is always the best of the four regardless of how many protected nodes there are.

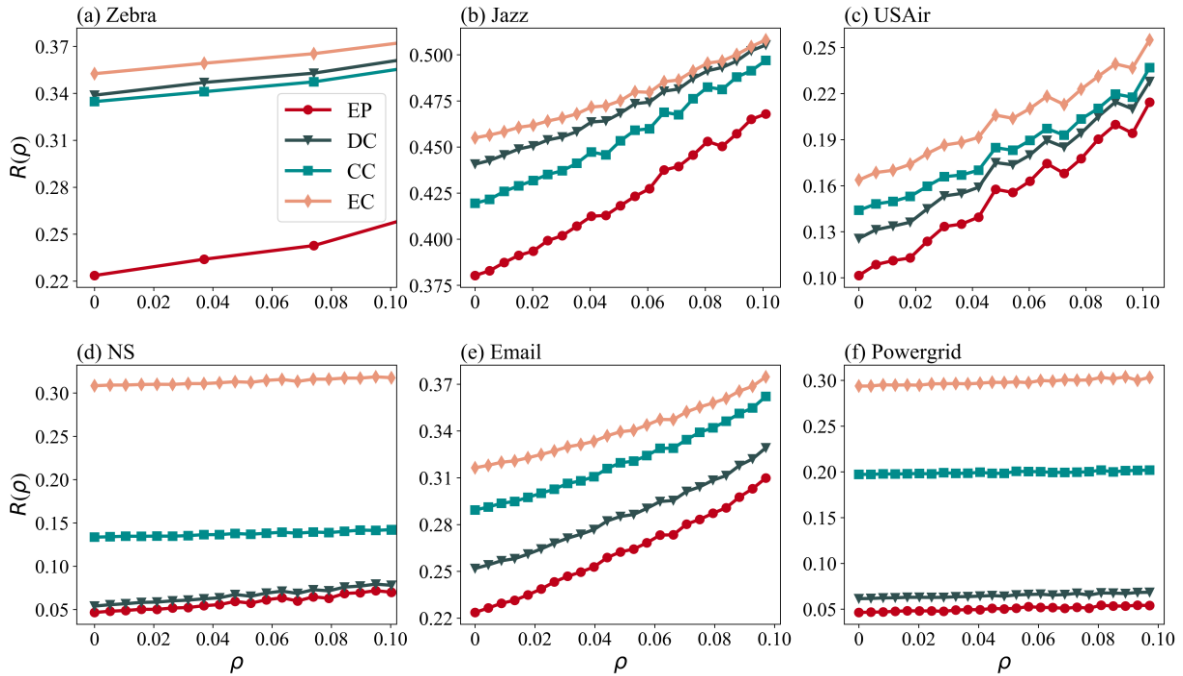

**Figure 4.** The influence of protection strategy on the performance of four indices.

*2.5.2 Noise tolerance.* Noisy data commonly exists in real networks, such as the addition of fake links or missing links [58,59]. To assess the tolerance of EP method to noisy data, for a given network $G$, we randomly remove or add $n$ ($n \in [0, 0.1N]$) links, and the resulted network is labeled as $G'$. Then the original network $G$ is attacked according to the rank which is calculated based on $G'$. The results are shown in Fig. 5. We can find that all four indices return similar responses to the noise. Specifically, the noise has the least influence on DC and the most influence on CC, while EC and EP are in between. But in all cases, EP method performs best.

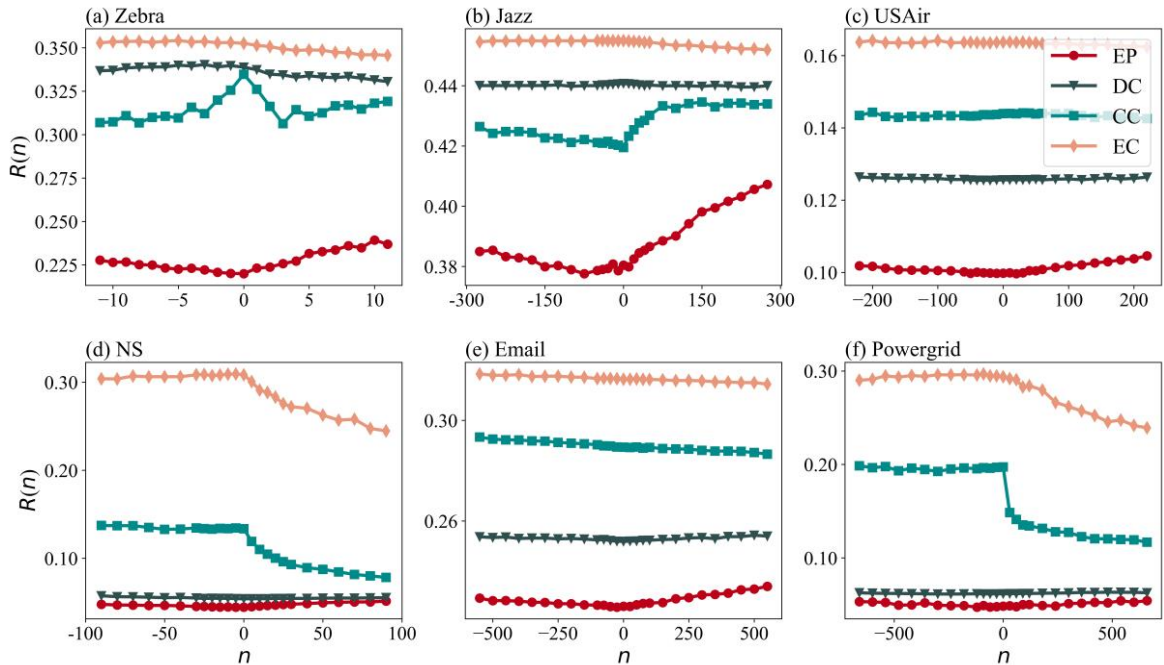

**Figure 5.** The influence of noise links on the performance of four indices. The positive and negative values of *n* correspond to the addition and removal of links, respectively. All the results are the average of 100 realizations.

Table 6. The Kendall correlation between four realizations of EP method on Powergrid.

| Realization | Robustness $R$ | $S_2$ | $S_3$ | $S_4$ |
|---|---|---|---|---|
| $S_1$ | 0.0458 | 0.753 | 0.752 | 0.757 |
| $S_2$ | 0.0464 |  | 0.755 | 0.756 |
| $S_3$ | 0.0464 |  |  | 0.758 |
| $S_4$ | 0.0480 |  |  |  |

## 2.6 The multi-solution advantages of EP method

Due to the randomness of the percolation process, the ranking obtained by EP method may differ in each realization, and thus may provide multiple optimal solutions. The nature of EP method is of great practical significance. For example, when some of the nodes are unable to be protected or attacked, the EP method can find out alternative rankings to achieve similar performance. Here, we take Powergrid network as an example to show this property.

We used the Kendall correlation coefficient $\tau$ (see Method section for details) to evaluate the difference of the rankings obtained from different realizations. The larger $\tau$, the more similar the two rankings are. The results presented in Table 6 suggest that the Robustness obtained from four different realizations($S_1$- $S_4$) are very close to each other (all fall in the range [0.0458, 0.0480]), which indicates that these rankings have similar attack effectiveness. While the values of $\tau$ between any two of them are not very high (0.752~0.758), indicating that there are still differences between the rankings obtained from different realizations. Therefore, the EP method allows us to get different rankings with almost the same performance, which can be used as a supplement to each other in practical applications. This property has also been used in the analysis of network robustness [60–62].

## 3. Discussion

In this paper, we proposed a vital node identification method based on Achlioptas process. Compared with classical methods, our method has better performance for networks with different structure features. This mainly because the information extracted from percolation clusters captures the local and global property of a network, simultaneously, so that the new method can well utilize the combined features to identify important nodes. Besides the high accuracy, the new method can also produce multiple optimal solutions due to the randomness of the percolation process. This advantage suggests that our method provides a variety of options for many practical problems, and thus has a wide range of applications, for instance, prevention and control of infectious diseases, power network maintenance, transportation or communication optimization, and so on.

For a given network $G(V, E)$, each realization of EP method can be regarded as a random node activation procedure with time complexity of $O(|V|)$. And the time complexity of activating a single node in the procedure, which includes the detection of cluster sizes using Breadth-first search, BFS [63], is $O(|V| + |E|)$. Thereofore, the time complexity of each realization is $O(|V|^2 + |V||E|)$. To achieve high accuracy, EP method needs to perform many realizations to reduce the fluctuation due to randomness, leading to high computational complexity, especially in very large scale networks. Fortunately, since each realization is independent, the method of parallel computation can thus be applied to improve the computational efficiency. In future work, we will further improve EP method, optimize the computational efficiency, and seek more practical application scenarios. Although the proposed method is to identify important nodes, this research idea and framework can inspire other problems in complex network research, such as link prediction, community decomposition, etc., which will be set as our future researches.

## Method

**Eigenvector centrality, EC** [2,15,37]. Eigenvector centrality assumes that the importance of a node depends not only on the number of its neighbors, but also on the importance of neighbors. It is defined as

$$\text{EC}_i = x_i = c \sum_{j=1}^n a_{ij} x_j \qquad (10)$$

where $x_i$ describes the importance of node $i$. For all the nodes, this equation can be written in a matrix form $\boldsymbol{x} = c\boldsymbol{A}\boldsymbol{x}$. Here, $c$ is a proportionality constant, defined as $c = 1/\lambda$ where $\lambda$ is the largest eigenvalue of the network adjacency matrix $\boldsymbol{A}$.

**Closeness centrality, CC** [35,36]. Closeness Centrality believes that the node has a smaller average distance with other nodes in the network is more important. Closeness centrality is defined as

$$\text{CC}_i = \frac{1}{n-1} \sum_{j \neq i} \frac{1}{d_{ij}} \qquad (11)$$

where $d_{ij}$ is the shortest path of network. If network is not connected, $d_{ij}$ may be $\infty$, and $1/\infty = 0$.

**Collective Influence, CI** [19,38]. Define $\text{Ball}(i, l)$ the set of nodes inside a ball of radius $l$ (defined as the shortest path) around node $i$, $\partial \text{Ball}(i, l)$ is the frontier of the ball. Then the CI strength of node $i$ at level $l$ is

$$\text{CI}_i = (k_i - 1) \sum_{j \in \partial \text{Ball}(i,l)} (k_j - 1) \qquad (12)$$

where $k_i$ is the degree of node $i$ and $l$ is a predefined nonnegative integer which does not exceed the network diameter for a finite network. In this paper, we set $l$ as 2.

**Adaptive Degree, AD** [38,39]. Adaptive degree is a variant of the degree method. AD recomputes the degree of all remaining nodes after each removal of the node with the highest degree in the current network.

**Kendall correlation coefficient** [64]. Considering two sequences $X, Y$ with $N$ elements, and $X_i$、$Y_i$ represent the $i$-th element in $X, Y$, respectively. If $i<j$ and $X_i - X_j$ has same sign with $Y_i - Y_j$, then we call pair $(i, j)$ consistent, and the formula of Kendall correlation coefficient is

$$\tau = \frac{2K}{n(n-1)} \qquad (12)$$

where

$$K = \sum_{i=1}^{n-1} \sum_{j=i+1}^{n} \varphi^*(X_i, X_j, Y_i, Y_j) \qquad (13)$$

$$\varphi^*(X_i, X_j, Y_i, Y_j) \begin{cases} 1 \ if \ (X_i - X_j)(Y_i - Y_j) > 0 \\ 0 \ if \ (X_i - X_j)(Y_i - Y_j) = 0 \\ -1 \ if \ (X_i - X_j)(Y_i - Y_j) < 0 \end{cases}. \qquad (14)$$

Obviously, $\tau \in [-1,1]$. $\tau = -1$ indicates that $X$ and $Y$ are two completely opposite ranks, while $\tau = 1$ means the two ranks are identical.

## Acknowledgements

This work is supported by the National Natural Science Foundation of China (Nos. 61673150, 11622538). Lü L. acknowledges the Science Strength Promotion Programme of UESTC, Chengdu and the Zhejiang Provincial Natural Science Foundation of China (No. LR16A050001).

## Author contributions

Zhihao Qiu, Ming Li, and Linyuan Lü designed the study. Zhihao Qiu performed the research and Tianlong Fan executed the experiment validation. All authors analyzed the data and discussed the results. Zhihao Qiu and Tianlong Fan drafted the manuscript. Ming Li and Linyuan Lü revised the manuscript and all authors read and approved the final manuscript.

## Conflict of interest

The authors declare that they have no conflict of interest.